\documentclass[aps,pra,twocolumn,amsmath,amssymb]{revtex4-1} 
\usepackage[utf8]{inputenc}
\usepackage[english]{babel}
\usepackage{graphicx}
\usepackage{subfiles}

\usepackage{ulem}
\usepackage{blindtext}
\usepackage{amssymb}
\usepackage{subcaption}
\usepackage{amsmath}
\usepackage{graphicx}
\usepackage[usenames,dvipsnames]{xcolor}
\usepackage[braket, qm]{qcircuit}
\usepackage{tikz}
\usepackage{float}

\usepackage{wrapfig}
\usepackage{xargs}
\usepackage[colorlinks]{hyperref}
\hypersetup{
    colorlinks=true,
    citecolor=blue,
    linkcolor=red,
    filecolor=red,
    urlcolor=blue,
}

\usepackage[margin=1in]{geometry}

\begin{document}

\title{Projective cooling for the transverse Ising model}
\author{Erik J. Gustafson$^1$}
\affiliation{$^1$ Department of Physics and Astronomy, The University of Iowa, Iowa City, IA 52242, USA}
\date{\today}
\begin{abstract}
We demonstrate the feasibility of ground state preparation for the transverse Ising model using projective cooling, and show that the algorithm can effectively construct the ground state in the disordered (paramagnetic) phase. On the other hand, significant temperature effects are encountered in the ordered (ferromagnetic) phase requiring larger lattices to accurately simulate. 
\end{abstract}
\maketitle
\section{Introduction}
\label{sec:introduction}
In quantum computing efficient state preparation methods are important for the analysis of quantum field theories so that field excitations can be accurately described. Much work has been done on algorithms to prepare ground states and excited states for interacting field theories. These algorithms include quantum adiabatic evolution \cite{Jordan2014QuantumAF,Jordan2011QuantumCO,Farhi2000QuantumCB}, variational methods \cite{PhysRevLett.120.210501,Peruzzo_2014,aless2019quantum}, quantum phase estimation \cite{kitaev1995quantum,PhysRevLett.79.2586}, more recently circumventing this problem using classically generated lattice configurations \cite{Harmalkar:2020mpd, PhysRevLett.121.170501}, and a recently proposed projective cooling algorithm \cite{lee2019projected}. The first four algorithms each have their strengths and weaknesses in the field of quantum computing that have been thoroughly examined; however not much work has focused on the strengths and weaknesses of projective cooling. 

The work done in \cite{lee2019projected} investigated models which conserved particle number. The authors demonstrated that their new algorithm is efficient in preparing bound states for these Hamiltonians. A natural extension is to examine a quantum field theory which has an effective ``pair" creation and annihilation, the transverse Ising model (TIM). This choice is inspired by the road map used to develop lattice computations for QCD \cite{RevModPhys.55.775,RevModPhys.51.659}, since it is a stepping stone toward understanding theories containing confinement or are strongly coupled. 

Sec. \ref{sec:theory} layouts out the projective cooling algorithm and the Hamiltonian that will be investigated. Sec. \ref{sec:results} shows the results for the asymptotic behavior for both the ordered and disordered phases, and finite size scaling behavior in the transverse Ising model. 

\begin{figure}[h]
    \includegraphics[width=0.5\textwidth]{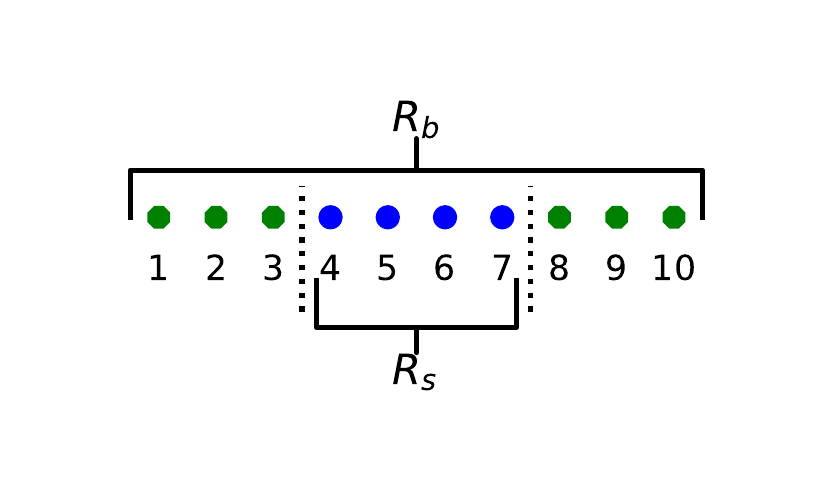}
    \vspace{-3em}
    \caption{Depiction of the regions $R_b$ and $R_s$ for a lattice with $N_b=10$ total sites and an $N_s=4$ sites contained within $R_s$. In this case $N_1 = 4$ and $N_2 = 7$}
    \label{fig:diagram}
\end{figure}
\section{Theory}
\label{sec:theory} 
The idea behind projective cooling involves removing high energy excitations outside of some region of interest by projecting them away. More explicitly, projective cooling works as follows (see Ref. \cite{lee2019projected} for more thorough details): a small region $R_{s}$, which contains $N_s$ sites and supports the Hamiltonian of interest $H_s$, is chosen so that it is symmetrically contained within some larger system $R_{b}$ (see Fig. \ref{fig:diagram}), which contains $N_b$ sites, with a corresponding Hamiltonian $\hat{H}_b$. An initial state $|\psi \rangle$ that has support on $R_s$ is prepared and time evolved corresponding to $\hat{H_b}$. This process is in may ways similar to the quantum Joule expansion \cite{Zhang:2019trs}. The difference arises when the particle excitations outside of the $R_s$ are projected away, and in the limit that $N_b \gg N_s$ the wave function in $R_s$ will approach an asymptotic state. This algorithm can be summarized algebraically, 
\begin{equation}
    \label{eq:projectivecooling}
    |\psi_0 \rangle = \mathcal{P} U(t) |\psi \rangle,
\end{equation}
where $\mathcal{P}$ is the operator that projects away excitations outside of the $R_s$, $U(t) = e^{-i t \hat{H}_b}$, and $|\psi\rangle$ is the initial wavefunction.

Two formulations for the TIM Hamiltonian were used in this work, depending on the quantum phase the system is in. The reason for choosing different formulations is a result of choosing a basis which is natural to work in. In the disordered phase ($J<h_T$), it is easier to work in a basis where the transverse field is diagonal, conversely, in the ordered phase ($J > h_T$) it is easier to work in a basis where the nearest neighbor coupling is diagonal. In the disordered phase, the formulation of the TIM Hamiltonian, in $R_s$, used in this work is
\begin{equation}
    \label{eq:disorderedham}
    \hat{H_s} = -J \sum_{i=N_1}^{N_2-1} \hat{\sigma}^x_i \hat{\sigma}^x_{i+1} - \sum_{i=N_1}^{N_2} \Big(h_T \hat{\sigma}^z_i + h \hat{\sigma}^x_i\Big)
\end{equation}
where $J$ is the nearest neighbor coupling, $h_T$ is the onsite energy, $h$ is the longitudinal field coupling which lifts the degeneracy in the strongly ordered phase ($h_T = 0$), $N_1 = (N_b - N_s)/2 + 1$, and $N_2 = (N_b + N_s)/2$. It should be noted that $N_b$ and $N_s$ must have the same parity. This choice of $N_1$ and $N_2$ forces $R_s$ to be symmetrically located within $R_b$. In this work $h_T = 1$, and $h=N_s^{-15/8}$ to ensure that the longitudinal field is perturbative. The choices of $h_T$ and $h$ are not arbitrary. Since the TIM undergoes a second order phase transition when $J=h_T$, setting $h_T=1$ has the phase transition occur at $J=1$. The choice of $h=N_s^{-15/8}$ is to ensure that finite size scaling relations are obeyed, and the effects of the longitudinal field are perturbative. \cite{Zhang:2019trs} showed that using this formulation for the TIM, that finite size scaling relations were obeyed even for small lattices of 8 sites. The Hamiltonian for $R_b$ in the disordered phase is
\begin{equation}
    \label{eq:nbham}
    \begin{split}
        \hat{H}_b &= \hat{H}_s - J \sum_{i=1}^{N_1-1} \Big(\hat{\sigma}^x_i\hat{\sigma}^x_{i+1} + \hat{\sigma}^y_i \hat{\sigma}^y_{i+1}\Big) \\
        & - J \sum_{i=N_2}^{N_b - 1} \Big(\hat{\sigma}^x_i\hat{\sigma}^x_{i+1} + \hat{\sigma}^y_i \hat{\sigma}^y_{i+1}\Big) - h_T \sum_{i \notin R_s} \hat{\sigma}^z_i \\
    \end{split}
\end{equation}
The different forms of the hopping terms inside and outside of $R_s$ are to ensure that the projected ground state outside of $R_s$ corresponds to all spins pointing up and the cooling only happens one way, away from $R_s$. 

The ordered phase the Hamiltonian is
\begin{equation}
\label{eq:orderedham}
\hat{H}_b = - J \sum_{i=1}^{N_b - 1} \hat{\sigma}^z_i \hat{\sigma}^z_{i+1} - \sum_{i=1}^{N_b} \Big(h_T \hat{\sigma}^x_i + N_b^{-15/8} \hat{\sigma}^z_i\Big).
\end{equation}
This Hamiltonian does not have a change in couplings is to ensure that the domain wall excitations do not bounce back into $R_s$; the risk of doing this is that there may be some ``heat" leaking from outside $R_s$, however later results will show that this is negligible. 

In all cases,  $N_b$ ranges from 6 to 14, $N_s$ ranges from 4 to 9 sites, and the initial state has all spins pointing up. Due to the size of some of the Hilbert spaces, the time evolution operator $U(t)$ is represented using a Suzuki-Trotter approximation; in the disordered phase the time evolution operator is
\begin{equation}
\begin{split}
    \hat{U}(t; \delta t)  \approx \Big(&
    e^{i \delta t J \sum_{i} \hat{\sigma}^y_i \hat{\sigma}^y_{i+1}}e^{i \delta t J \sum_{i} \hat{\sigma}^x_i \hat{\sigma}^x_{i+1}} \\
    &e^{i \delta t h_T \sum_i \hat{\sigma}^z_i} e^{i h \delta t \sum_i \hat{\sigma}^x_i} \Big)^{\frac{t}{\delta t}},
\end{split}
\end{equation} 
while in the ordered phase the time evolution operator is 
\begin{equation}
\begin{split}
    \hat{U}(t; \delta t)  \approx \Big(& e^{i \delta t J \sum_{i} \hat{\sigma}^z_i \hat{\sigma}^z_{i+1}} e^{i \delta t h_T \sum_i \hat{\sigma}^x_i} e^{i h \delta t \sum_i \hat{\sigma}^z_i} \Big)^{\frac{t}{\delta t}}
\end{split}
\end{equation}
In all cases $\delta t = 0.01 / J$ so as to keep the systematic error from this approximation negligible for large time scales.
\section{Results}
\label{sec:results}
It is important to first ensure that the system will approach a stable asymptotic state. This can be done by measuring the overlap of the time evolved and projected state with the actual ground state of the Hamiltonian, as done in \cite{lee2019projected}; however, on a quantum computer it is not possible to measure this overlap. In keeping with the methods of quantum computing we can expect to see asymptotic behavior by measuring the energy density of the projected state in the compact region $R_s$. 

\begin{figure}
    \includegraphics[width=0.48\textwidth]{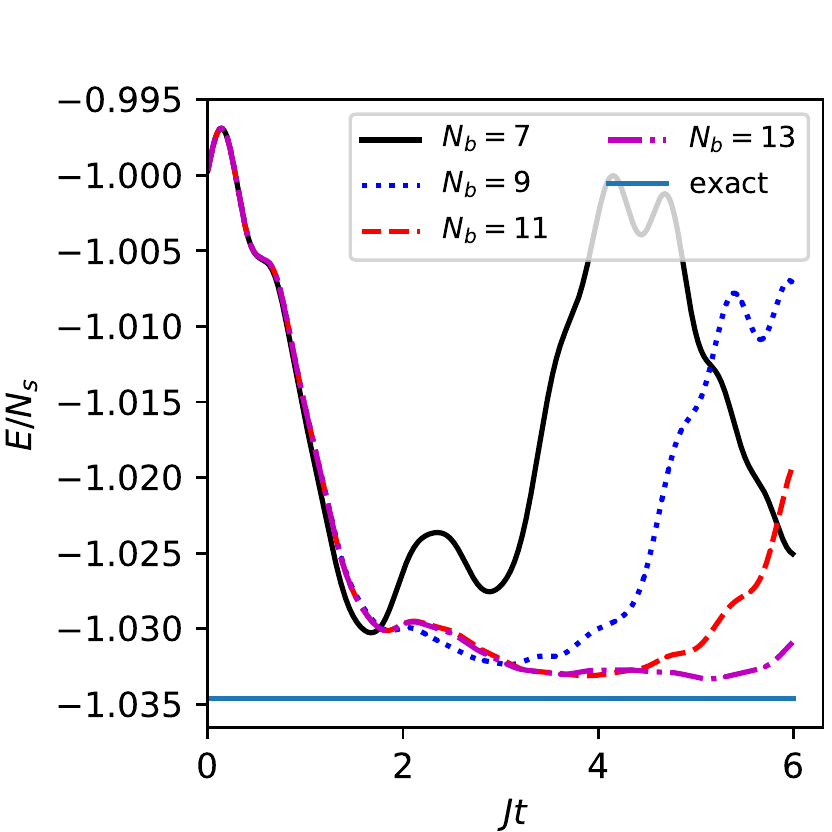}
    \caption{Energy per site (in the small region) of projectively cooled state as a function of time in the ordered phase; $J=0.4$, $N_s=5$}
    \label{fig:disorderfixedpoint1}
\end{figure}

\begin{figure}[ht]
    \includegraphics[width=0.5\textwidth]{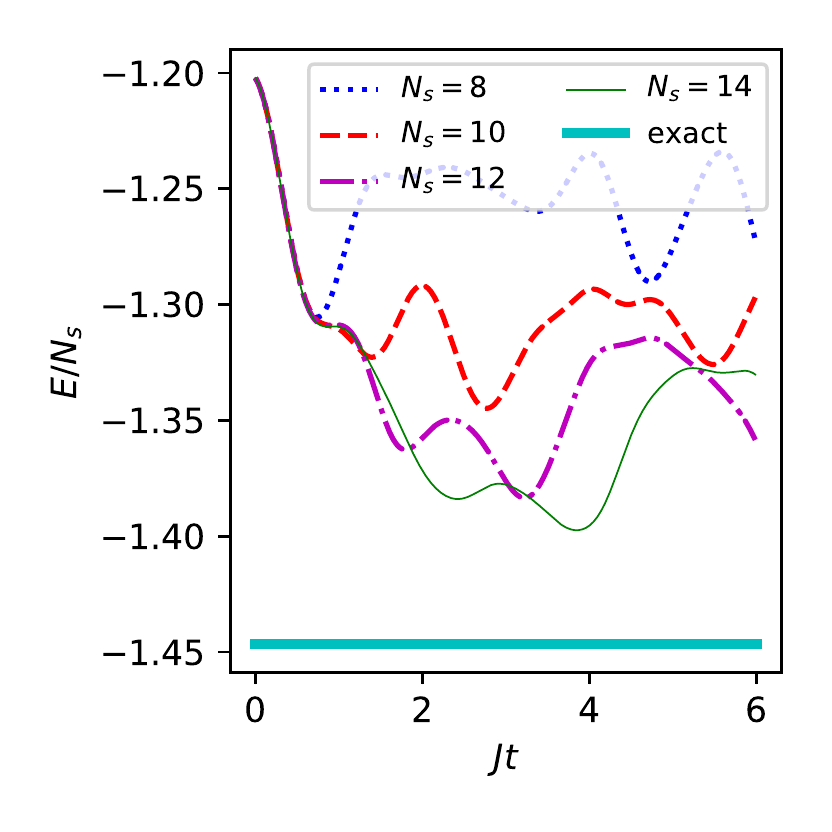}
    \caption{Energy per site of the state using projective cooling as a function of time in the ordered phase; $J=1.4$, $n_s=6$.}
    \label{fig:orderedfixedpoint1}
\end{figure}

Fig. \ref{fig:disorderfixedpoint1}
shows a typical result ($J=0.4$ and $N_s=5$) for the disordered phase. It is clear that the system approaches an approximate plateau for $N_b=11,$ and $13$ and less so for $N_b=9$ and does not approach a plateau at all for $N_b=7$. Fig. \ref{fig:orderedfixedpoint1} demonstrates the evolution toward an asymptotic state in the ordered phase for $J=1.4$ and $N_s = 6$.
\begin{figure}[ht]
    \centering
    \includegraphics[width=0.5\textwidth]{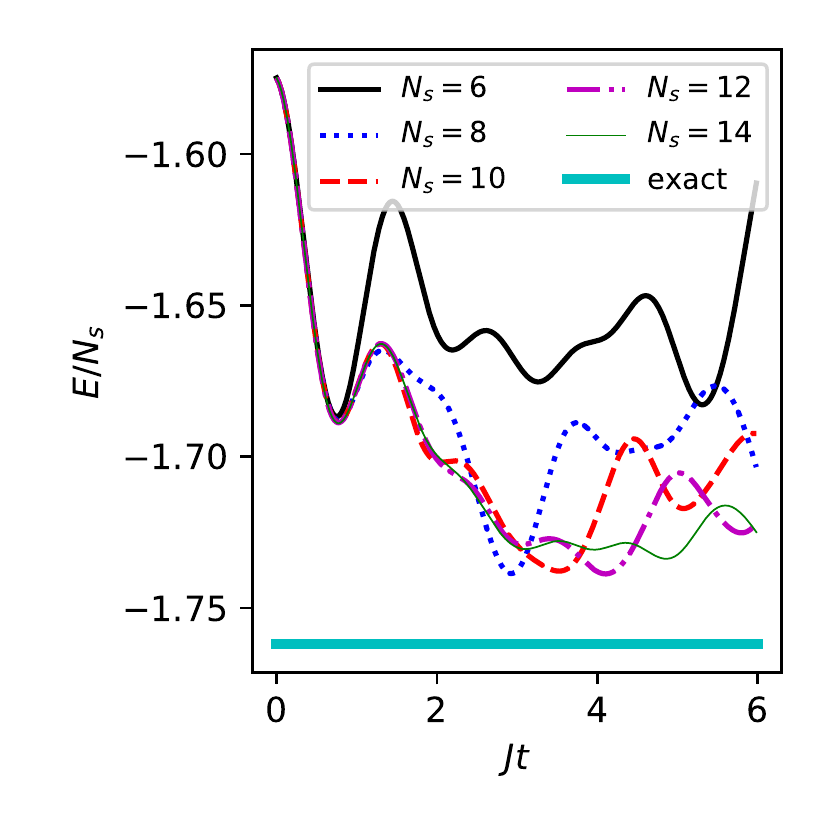}
    \caption{Energy per site of the state using projective cooling as a function of time in the ordered phase; $J=2.0$, $n_s=4$.}
    \label{fig:orderedfixedpoint2}
\end{figure}
The results of the ordered regime are more noisy because there is heat leaking back into $R_s$ \footnote{Attempting to use an eigenstate in the $\sigma^x$ basis for $R_s$ and time evolve according to Hamiltonian in Eq. \ref{eq:nbham} for the ordered phase yielded worse results because excitations were immediately reflected back into $R_s$.}. The noticeable and important feature that arises is as $N_b$ increases, the minimum of the energy density approaches the exact value. This is reassuring even if we do not see the same plateau. In Fig. \ref{fig:orderedfixedpoint2} ($J=2.0$ and $N_s=4$), we see the plateaus become more noticeable again as $N_b$ increases but they are not as clean as the plateaus in the disordered phase.

In Figs. \ref{fig:disorderfixedpoint1}, \ref{fig:orderedfixedpoint1}, and \ref{fig:orderedfixedpoint2} the energy density of the asymptotic state, in general, becomes closer to the exact energy density as $N_b$ becomes larger. This is indicative that $N_b$ introduces lattice artifacts to the calculation because it is finite. These artifacts can be mitigated by extrapolating to the limit where the volume of $R_b$ is infinite. In order to do this, the following ansatz was chosen,
\begin{equation}
    \label{eq:infinitevolume}
    E(N_b) = A e^{-B N_b} + E_{\infty}.
\end{equation}
This ansatz constrains the energy density to always be finite and approach an asymptotic value as $N_b \rightarrow \infty$.

\begin{figure}
    \centering
    \includegraphics[width=0.5\textwidth]{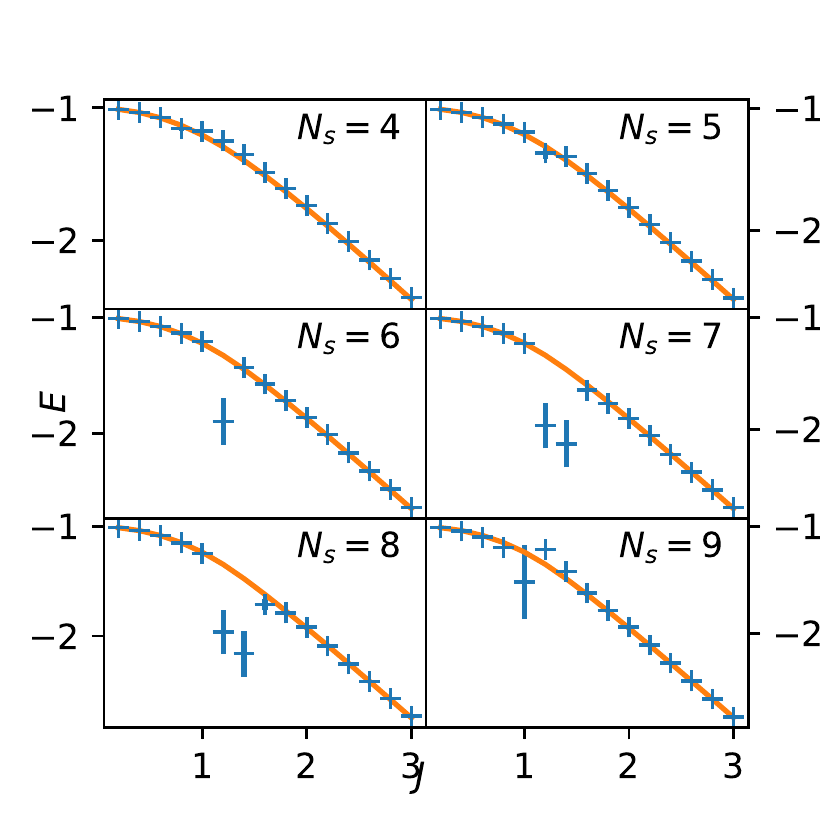}
    \caption{Energy density using an infinite volume extrapolation as a function of the coupling constant for various compact region sizes. Blue (online) crosses: extrapolated points; orange (online) curve: ground state energy density via exact diagonalization.}
    \label{fig:extrapolation}
    \centering
    \includegraphics[width=0.48\textwidth]{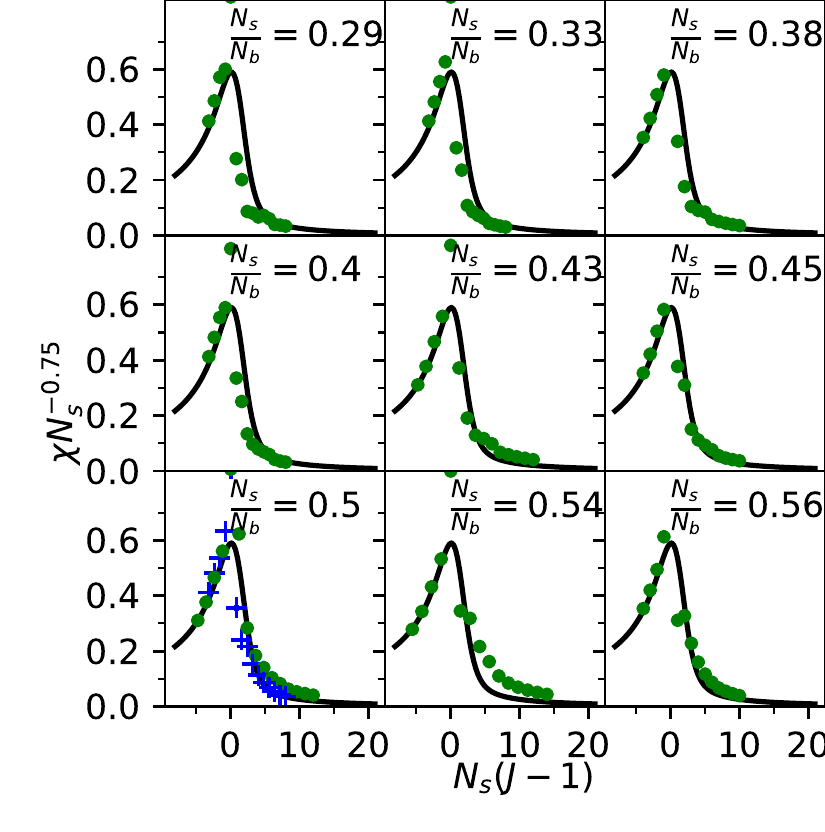}
    \caption{Re-scaled magnetic susceptibility as a function of the re-scaled nearest neighbor coupling for various ratios of $N_s/N_b$. Black (online) curve: is an interpolation for the exact magnetic susceptibility for 14 sites; green (online) points and blue (online) crosses: the calculated susceptibilities using projective cooling.}
    \label{fig:datacollapse}
\end{figure}
The range from 2 per-cent above the minimal value of the energy density (traced back from the minimal value) to the minimum of the energy density to determine the corresponding uncertainty for the energy density. The result of this algorithm favors simulations that exhibit a plateau verses an local minimum. Fig. \ref{fig:extrapolation} shows the infinite volume extrapolation. There is excellent overlap with exact results away from $J=1$, but does show some substantial deviation near the phase transition when $N_s > 5$.

A second feature that is indicative the system is close to the ground state is the finite size scaling relations for the magnetic susceptibility are preserved. The susceptiblity is defined as
\begin{equation}
    \chi = \frac{1}{N_s} 
    \begin{cases}
    \sum_{i,j=1}^{N_s} \langle \hat{\sigma}^x_i \hat{\sigma}^x_j \rangle - \langle \hat{\sigma}^x_i \rangle \langle \hat{\sigma}^x_j \rangle & J < h_T \\
    \sum_{i,j=1}^{N_s} \langle \hat{\sigma}^z_i \hat{\sigma}^z_j \rangle - \langle \hat{\sigma}^z_i \rangle \langle \hat{\sigma}^z_j \rangle & J \geq h_T \\
    \end{cases},
\end{equation}
where the different formulas correspond to the different bases that are worked in. The susceptibilities are calculated over the same region that the energy densities are. The data collapse is demonstrated in Fig. \ref{fig:datacollapse} where different ratios of $N_s/N_b$ are plotted to demonstrate possible thermal effects. For ratios of $N_s/N_b > 5/9$ non-linear effects begin to take over and cause the finite size scaling to break down and are not shown. 
\section{Conclusions}
\label{sec:conclusions}
This work demonstrates that projective cooling can effectively and accurately prepare the ground state for a relatively simple field theory with a non-trivial ground state. The projective cooling algorithm constructs the ground state in the disordered phase of the transverse Ising model more accurately than in the ordered phase. The discrepancies in the ordered phase are likely a result of thermal effects, indicated by the noticeable discrepancies of the magnetic susceptibility in the ordered phase.

The work done here can be extended to extracting bound states energies for attractive interacting problems such as an Ising-like model with both $\hat{\sigma}^z\hat{\sigma}^z$ and $\hat{\sigma}^x\hat{\sigma}^x$ interactions with only a few changes to the choice of initial state. Other possible extensions could be the Schwinger or O(N) models. In addition, optimizing this algorithm for quantum computation is a challenge that must be addressed as the readout errors and machine noise outside of the $R_s$ can have a drastic effect on the interpreted states, and the costs of post-selection using the projection operator.
\begin{acknowledgments}
This work was supported in part by the U.S. Department of Energy (DOE) under Award Number DE-SC0019139. I thank Yannick Meurice, Wayne Polyzou, Dean Lee, and Henry Lamm for fruitful conversations. 
\end{acknowledgments}
\vspace*{\fill}
\bibliographystyle{apsrev4-1}
\bibliography{main.bbl}

\begin{thebibliography}{15}%
\makeatletter
\providecommand \@ifxundefined [1]{%
 \@ifx{#1\undefined}
}%
\providecommand \@ifnum [1]{%
 \ifnum #1\expandafter \@firstoftwo
 \else \expandafter \@secondoftwo
 \fi
}%
\providecommand \@ifx [1]{%
 \ifx #1\expandafter \@firstoftwo
 \else \expandafter \@secondoftwo
 \fi
}%
\providecommand \natexlab [1]{#1}%
\providecommand \enquote  [1]{``#1''}%
\providecommand \bibnamefont  [1]{#1}%
\providecommand \bibfnamefont [1]{#1}%
\providecommand \citenamefont [1]{#1}%
\providecommand \href@noop [0]{\@secondoftwo}%
\providecommand \href [0]{\begingroup \@sanitize@url \@href}%
\providecommand \@href[1]{\@@startlink{#1}\@@href}%
\providecommand \@@href[1]{\endgroup#1\@@endlink}%
\providecommand \@sanitize@url [0]{\catcode `\\12\catcode `\$12\catcode
  `\&12\catcode `\#12\catcode `\^12\catcode `\_12\catcode `\%12\relax}%
\providecommand \@@startlink[1]{}%
\providecommand \@@endlink[0]{}%
\providecommand \url  [0]{\begingroup\@sanitize@url \@url }%
\providecommand \@url [1]{\endgroup\@href {#1}{\urlprefix }}%
\providecommand \urlprefix  [0]{URL }%
\providecommand \Eprint [0]{\href }%
\providecommand \doibase [0]{http://dx.doi.org/}%
\providecommand \selectlanguage [0]{\@gobble}%
\providecommand \bibinfo  [0]{\@secondoftwo}%
\providecommand \bibfield  [0]{\@secondoftwo}%
\providecommand \translation [1]{[#1]}%
\providecommand \BibitemOpen [0]{}%
\providecommand \bibitemStop [0]{}%
\providecommand \bibitemNoStop [0]{.\EOS\space}%
\providecommand \EOS [0]{\spacefactor3000\relax}%
\providecommand \BibitemShut  [1]{\csname bibitem#1\endcsname}%
\let\auto@bib@innerbib\@empty
\bibitem [{\citenamefont {Jordan}\ \emph {et~al.}(2014)\citenamefont {Jordan},
  \citenamefont {Lee},\ and\ \citenamefont {Preskill}}]{Jordan2014QuantumAF}%
  \BibitemOpen
  \bibfield  {author} {\bibinfo {author} {\bibfnamefont {S.~P.}\ \bibnamefont
  {Jordan}}, \bibinfo {author} {\bibfnamefont {K.~S.~M.}\ \bibnamefont {Lee}},
  \ and\ \bibinfo {author} {\bibfnamefont {J.}~\bibnamefont {Preskill}}\
  }(\bibinfo {year} {2014})\BibitemShut {NoStop}%
\bibitem [{\citenamefont {Jordan}\ \emph {et~al.}(2011)\citenamefont {Jordan},
  \citenamefont {Lee},\ and\ \citenamefont {Preskill}}]{Jordan2011QuantumCO}%
  \BibitemOpen
  \bibfield  {author} {\bibinfo {author} {\bibfnamefont {S.~P.}\ \bibnamefont
  {Jordan}}, \bibinfo {author} {\bibfnamefont {K.~S.~M.}\ \bibnamefont {Lee}},
  \ and\ \bibinfo {author} {\bibfnamefont {J.}~\bibnamefont {Preskill}},\
  }\href@noop {} {\bibfield  {journal} {\bibinfo  {journal} {Quantum
  Information and Computation}\ }\textbf {\bibinfo {volume} {14}},\ \bibinfo
  {pages} {1014} (\bibinfo {year} {2011})}\BibitemShut {NoStop}%
\bibitem [{\citenamefont {Farhi}\ \emph {et~al.}(2000)\citenamefont {Farhi},
  \citenamefont {Goldstone}, \citenamefont {Gutmann},\ and\ \citenamefont
  {Sipser}}]{Farhi2000QuantumCB}%
  \BibitemOpen
  \bibfield  {author} {\bibinfo {author} {\bibfnamefont {E.}~\bibnamefont
  {Farhi}}, \bibinfo {author} {\bibfnamefont {J.}~\bibnamefont {Goldstone}},
  \bibinfo {author} {\bibfnamefont {S.}~\bibnamefont {Gutmann}}, \ and\
  \bibinfo {author} {\bibfnamefont {M.}~\bibnamefont {Sipser}}\ }(\bibinfo
  {year} {2000})\BibitemShut {NoStop}%
\bibitem [{\citenamefont {Dumitrescu}\ \emph {et~al.}(2018)\citenamefont
  {Dumitrescu}, \citenamefont {McCaskey}, \citenamefont {Hagen}, \citenamefont
  {Jansen}, \citenamefont {Morris}, \citenamefont {Papenbrock}, \citenamefont
  {Pooser}, \citenamefont {Dean},\ and\ \citenamefont
  {Lougovski}}]{PhysRevLett.120.210501}%
  \BibitemOpen
  \bibfield  {author} {\bibinfo {author} {\bibfnamefont {E.~F.}\ \bibnamefont
  {Dumitrescu}}, \bibinfo {author} {\bibfnamefont {A.~J.}\ \bibnamefont
  {McCaskey}}, \bibinfo {author} {\bibfnamefont {G.}~\bibnamefont {Hagen}},
  \bibinfo {author} {\bibfnamefont {G.~R.}\ \bibnamefont {Jansen}}, \bibinfo
  {author} {\bibfnamefont {T.~D.}\ \bibnamefont {Morris}}, \bibinfo {author}
  {\bibfnamefont {T.}~\bibnamefont {Papenbrock}}, \bibinfo {author}
  {\bibfnamefont {R.~C.}\ \bibnamefont {Pooser}}, \bibinfo {author}
  {\bibfnamefont {D.~J.}\ \bibnamefont {Dean}}, \ and\ \bibinfo {author}
  {\bibfnamefont {P.}~\bibnamefont {Lougovski}},\ }\href {\doibase
  10.1103/PhysRevLett.120.210501} {\bibfield  {journal} {\bibinfo  {journal}
  {Phys. Rev. Lett.}\ }\textbf {\bibinfo {volume} {120}},\ \bibinfo {pages}
  {210501} (\bibinfo {year} {2018})}\BibitemShut {NoStop}%
\bibitem [{\citenamefont {Peruzzo}\ \emph {et~al.}(2014)\citenamefont
  {Peruzzo}, \citenamefont {McClean}, \citenamefont {Shadbolt}, \citenamefont
  {Yung}, \citenamefont {Zhou}, \citenamefont {Love}, \citenamefont
  {Aspuru-Guzik},\ and\ \citenamefont {O’Brien}}]{Peruzzo_2014}%
  \BibitemOpen
  \bibfield  {author} {\bibinfo {author} {\bibfnamefont {A.}~\bibnamefont
  {Peruzzo}}, \bibinfo {author} {\bibfnamefont {J.}~\bibnamefont {McClean}},
  \bibinfo {author} {\bibfnamefont {P.}~\bibnamefont {Shadbolt}}, \bibinfo
  {author} {\bibfnamefont {M.-H.}\ \bibnamefont {Yung}}, \bibinfo {author}
  {\bibfnamefont {X.-Q.}\ \bibnamefont {Zhou}}, \bibinfo {author}
  {\bibfnamefont {P.~J.}\ \bibnamefont {Love}}, \bibinfo {author}
  {\bibfnamefont {A.}~\bibnamefont {Aspuru-Guzik}}, \ and\ \bibinfo {author}
  {\bibfnamefont {J.~L.}\ \bibnamefont {O’Brien}},\ }\href {\doibase
  10.1038/ncomms5213} {\bibfield  {journal} {\bibinfo  {journal} {Nature
  Communications}\ }\textbf {\bibinfo {volume} {5}} (\bibinfo {year} {2014}),\
  10.1038/ncomms5213}\BibitemShut {NoStop}%
\bibitem [{\citenamefont {Roggero}\ \emph {et~al.}(2019)\citenamefont
  {Roggero}, \citenamefont {Li}, \citenamefont {Carlson}, \citenamefont
  {Gupta},\ and\ \citenamefont {Perdue}}]{aless2019quantum}%
  \BibitemOpen
  \bibfield  {author} {\bibinfo {author} {\bibfnamefont {A.}~\bibnamefont
  {Roggero}}, \bibinfo {author} {\bibfnamefont {A.~C.~Y.}\ \bibnamefont {Li}},
  \bibinfo {author} {\bibfnamefont {J.}~\bibnamefont {Carlson}}, \bibinfo
  {author} {\bibfnamefont {R.}~\bibnamefont {Gupta}}, \ and\ \bibinfo {author}
  {\bibfnamefont {G.~N.}\ \bibnamefont {Perdue}},\ }\href@noop {} {\enquote
  {\bibinfo {title} {Quantum computing for neutrino-nucleus scattering},}\ }
  (\bibinfo {year} {2019}),\ \Eprint {http://arxiv.org/abs/1911.06368}
  {arXiv:1911.06368 [quant-ph]} \BibitemShut {NoStop}%
\bibitem [{\citenamefont {Kitaev}(1995)}]{kitaev1995quantum}%
  \BibitemOpen
  \bibfield  {author} {\bibinfo {author} {\bibfnamefont {A.~Y.}\ \bibnamefont
  {Kitaev}},\ }\href@noop {} {\enquote {\bibinfo {title} {Quantum measurements
  and the abelian stabilizer problem},}\ } (\bibinfo {year} {1995}),\ \Eprint
  {http://arxiv.org/abs/quant-ph/9511026} {arXiv:quant-ph/9511026 [quant-ph]}
  \BibitemShut {NoStop}%
\bibitem [{\citenamefont {Abrams}\ and\ \citenamefont
  {Lloyd}(1997)}]{PhysRevLett.79.2586}%
  \BibitemOpen
  \bibfield  {author} {\bibinfo {author} {\bibfnamefont {D.~S.}\ \bibnamefont
  {Abrams}}\ and\ \bibinfo {author} {\bibfnamefont {S.}~\bibnamefont {Lloyd}},\
  }\href {\doibase 10.1103/PhysRevLett.79.2586} {\bibfield  {journal} {\bibinfo
   {journal} {Phys. Rev. Lett.}\ }\textbf {\bibinfo {volume} {79}},\ \bibinfo
  {pages} {2586} (\bibinfo {year} {1997})}\BibitemShut {NoStop}%
\bibitem [{\citenamefont {Harmalkar}\ \emph {et~al.}(2020)\citenamefont
  {Harmalkar}, \citenamefont {Lamm},\ and\ \citenamefont
  {Lawrence}}]{Harmalkar:2020mpd}%
  \BibitemOpen
  \bibfield  {author} {\bibinfo {author} {\bibfnamefont {S.}~\bibnamefont
  {Harmalkar}}, \bibinfo {author} {\bibfnamefont {H.}~\bibnamefont {Lamm}}, \
  and\ \bibinfo {author} {\bibfnamefont {S.}~\bibnamefont {Lawrence}},\
  }\href@noop {} {\  (\bibinfo {year} {2020})},\ \Eprint
  {http://arxiv.org/abs/2001.11490} {arXiv:2001.11490 [hep-lat]} \BibitemShut
  {NoStop}%
\bibitem [{\citenamefont {Lamm}\ and\ \citenamefont
  {Lawrence}(2018)}]{PhysRevLett.121.170501}%
  \BibitemOpen
  \bibfield  {author} {\bibinfo {author} {\bibfnamefont {H.}~\bibnamefont
  {Lamm}}\ and\ \bibinfo {author} {\bibfnamefont {S.}~\bibnamefont
  {Lawrence}},\ }\href {\doibase 10.1103/PhysRevLett.121.170501} {\bibfield
  {journal} {\bibinfo  {journal} {Phys. Rev. Lett.}\ }\textbf {\bibinfo
  {volume} {121}},\ \bibinfo {pages} {170501} (\bibinfo {year}
  {2018})}\BibitemShut {NoStop}%
\bibitem [{\citenamefont {Lee}\ \emph {et~al.}(2019)\citenamefont {Lee},
  \citenamefont {Bonitati}, \citenamefont {Given}, \citenamefont {Hicks},
  \citenamefont {Li}, \citenamefont {Lu}, \citenamefont {Rai}, \citenamefont
  {Sarkar},\ and\ \citenamefont {Watkins}}]{lee2019projected}%
  \BibitemOpen
  \bibfield  {author} {\bibinfo {author} {\bibfnamefont {D.}~\bibnamefont
  {Lee}}, \bibinfo {author} {\bibfnamefont {J.}~\bibnamefont {Bonitati}},
  \bibinfo {author} {\bibfnamefont {G.}~\bibnamefont {Given}}, \bibinfo
  {author} {\bibfnamefont {C.}~\bibnamefont {Hicks}}, \bibinfo {author}
  {\bibfnamefont {N.}~\bibnamefont {Li}}, \bibinfo {author} {\bibfnamefont
  {B.-N.}\ \bibnamefont {Lu}}, \bibinfo {author} {\bibfnamefont
  {A.}~\bibnamefont {Rai}}, \bibinfo {author} {\bibfnamefont {A.}~\bibnamefont
  {Sarkar}}, \ and\ \bibinfo {author} {\bibfnamefont {J.}~\bibnamefont
  {Watkins}},\ }\href@noop {} {\enquote {\bibinfo {title} {Projected cooling
  algorithm for quantum computation},}\ } (\bibinfo {year} {2019}),\ \Eprint
  {http://arxiv.org/abs/1910.07708} {arXiv:1910.07708 [quant-ph]} \BibitemShut
  {NoStop}%
\bibitem [{\citenamefont {Kogut}(1983)}]{RevModPhys.55.775}%
  \BibitemOpen
  \bibfield  {author} {\bibinfo {author} {\bibfnamefont {J.~B.}\ \bibnamefont
  {Kogut}},\ }\href {\doibase 10.1103/RevModPhys.55.775} {\bibfield  {journal}
  {\bibinfo  {journal} {Rev. Mod. Phys.}\ }\textbf {\bibinfo {volume} {55}},\
  \bibinfo {pages} {775} (\bibinfo {year} {1983})}\BibitemShut {NoStop}%
\bibitem [{\citenamefont {Kogut}(1979)}]{RevModPhys.51.659}%
  \BibitemOpen
  \bibfield  {author} {\bibinfo {author} {\bibfnamefont {J.~B.}\ \bibnamefont
  {Kogut}},\ }\href {\doibase 10.1103/RevModPhys.51.659} {\bibfield  {journal}
  {\bibinfo  {journal} {Rev. Mod. Phys.}\ }\textbf {\bibinfo {volume} {51}},\
  \bibinfo {pages} {659} (\bibinfo {year} {1979})}\BibitemShut {NoStop}%
\bibitem [{\citenamefont {Zhang}\ \emph {et~al.}(2019)\citenamefont {Zhang},
  \citenamefont {Meurice},\ and\ \citenamefont {Tsai}}]{Zhang:2019trs}%
  \BibitemOpen
  \bibfield  {author} {\bibinfo {author} {\bibfnamefont {J.}~\bibnamefont
  {Zhang}}, \bibinfo {author} {\bibfnamefont {Y.}~\bibnamefont {Meurice}}, \
  and\ \bibinfo {author} {\bibfnamefont {S.~W.}\ \bibnamefont {Tsai}},\
  }\href@noop {} {\  (\bibinfo {year} {2019})},\ \Eprint
  {http://arxiv.org/abs/1903.01414} {arXiv:1903.01414 [cond-mat.quant-gas]}
  \BibitemShut {NoStop}%
\bibitem [{Note1()}]{Note1}%
  \BibitemOpen
  \bibinfo {note} {Attempting to use an eigenstate in the $\sigma ^x$ basis for
  $R_s$ and time evolve according to Hamiltonian in Eq. \ref {eq:nbham} for the
  ordered phase yielded worse results because excitations were immediately
  reflected back into $R_s$.}\BibitemShut {Stop}%
\end{thebibliography}%

\newpage
\end{document}